\documentclass[%
reprint,
longbibliography,
amsmath,amssymb,
aps,
]{revtex4-1}

\usepackage{graphicx}
\usepackage{dcolumn}
\usepackage{bm}
\usepackage{xcolor, soul}
\usepackage{upgreek}
\usepackage{appendix}
\usepackage{float}
\usepackage{natbib}
\begin{document}

\preprint{APS/123-QED}

\title{The effect of bias current configuration on the performance of SQUID arrays}

\author{M. A. Gal\'i Labarias}
\email{marc.galilabarias@csiro.au}
\author{K.-H. M\"uller}
\author{E. E. Mitchell}%
\affiliation{%
 CSIRO Manufacturing, Lindfield, NSW, Australia.
}%




\date{\today}

\begin{abstract}
 Designing superconducting electronic devices involves a careful study of all the elements in the circuit, including the superconducting bias leads, however, the effect of the bias current leads on the performance of the device have not been previously addressed. 
In this work, we introduce a theoretical model for two-dimensional superconducting quantum interference device (SQUID) arrays capable of simulating the response of devices with different bias current configurations. 
First, we compare uniformly biased and centre biased SQUID arrays by investigating the voltage versus magnetic flux response, maximum transfer function and voltage modulation depth. 
Then, we calculate the time-averaged fluxoid distributions for 1D and 2D centre biased arrays.
Finally, we study the performance of the two bias current configurations depending on array size, screening parameter, thermal noise strength and kinetic self-inductance fraction.
Our calculations reveal: (i) centre biased 1D parallel SQUID arrays present an unusual voltage response caused by the asymmetric fluxoid distribution;
(ii) the optimal transfer function of centre biased arrays strongly depends on the number of junctions in parallel;
(iii) the performance of centre biased arrays approaches the uniform biased ones when the number of junctions in series exceeds those in parallel;
(iv) while the screening parameter and the thermal noise strength clearly affect the device performance, the kinetic self-inductance fraction seems to play only a secondary role.


\end{abstract}

\maketitle

\section{\label{sec:Intro}Introduction}

Arrays of superconducting quantum interference devices (SQUIDs) have been studied as a means of improving the magnetic field sensitivity \cite{Welty1991}, noise performance \cite{Krey1999}, dynamic range \cite{Foglietti1993, Kornev2009b}, linearity \cite{Kornev2009a} and robustness \cite{Oppenlander2000} over that of a single dc-SQUID \cite{Tesche1977, Clarke2004}. 
The performance of SQUID arrays depends on a range of parameters, including the device geometry \cite{Tolpygo1996, Muller2021} and superconducting film properties \cite{Ruffieux2020, Keenan2021}, the number of Josephson junctions (JJs) \cite{Oppenlander2000, Oppenlander2003, Kornev2009a} and their distribution throughout the array \cite{Mitchell2019, Gali2022a} and operating temperature \cite{Tesche1977, Taylor2016, Muller2021, Gali2022a}. 

Theoretical research has focused on design optimization of one-dimensional (1D) and two-dimensional (2D) SQUID arrays and superconducting quantum interference filters (SQIFs) to improve sensitivity, dynamic range and robustness. The initial studies relied on assumptions such as the absence of thermal noise  or a negligible screening parameter \cite{Oppenlander2000, Oppenlander2003, Kornev2009, Kornev2011}.
More recent models included inductance effects but thermal noise was still omitted \cite{Dalichaouch2014}.
Despite these methods being relevant when describing trends in low temperature superconducting (LTS) arrays, they are insufficient when trying to replicate experimental data for high-temperature superconducting (HTS) arrays \cite{Mitchell2016}. Experimental results of 1D HTS SQUID arrays showed unexpected features including deviations from the expected sinusoidal-like response and a decrease of the magnetic field sensitivity with the array width \cite{Mitchell2019} while 2D SQIF arrays showed much lower magnetic field sensitivity per loop than expected when compared with a single dc-SQUID \cite{Mitchell2016, Cybart2012}. 
Recent theoretical modelling of 1D parallel SQUID arrays has shown remarkably good agreement with experimental data \cite{Muller2021}.
This model demonstrated that to accurately replicate the experimental results of HTS SQUID arrays, it is necessary to consider not only the geometry of the individual SQUID loops, but the geometry of all the superconducting elements surrounding and connecting the  array. 
Recently, using a theoretical model, it has also been shown that the  inclusion of thermal noise at the operating temperature is necessary to predict the optimum bias current of 2D SQUID and SQIF arrays when accurately modelling their transfer function \cite{Gali2022a}.
Another gap between theory and experiment is that many models assume uniform biased arrays \cite{Oppenlander2000, Dalichaouch2014}.
However experimental devices are often designed with single bias leads to and from the array \cite{Gerdemann1995, Mitchell2016, Taylor2016, Mitchell2019, Muller2021} to make efficient use of the limited substrate surface area, limit trapping of vortices near the device and reduce circuit complexity. 
As device structure increases in size, issues such as multiple bias lines, multiplexing and heat-load limitations become important device design features.
However, as shown recently \cite{Mitchell2019}, the bias configuration choice can greatly affect the device performance, especially for arrays with many junctions connected in parallel. 

In this work we theoretically compare two bias current configurations: arrays with a single centred bias lead (centre bias) and arrays with a bias lead per junction in parallel (uniform bias). 
We study the flux-to-voltage response and transfer function of 1D and 2D devices including thermal noise and biased at the optimal current and applied magnetic flux. 
For  both configurations, we show the dependence of the optimal transfer function on the number of junctions in series ($N_s$) and in parallel ($N_p$) for different screening parameters, thermal noise strengths and kinetic inductance fractions and discuss the appearance of large fluxoids when the centre bias configuration is used.
In particular, our model helps explain some of the unusual features observed experimentally.


\section{Results and discussion}

Experimental SQUID arrays often use single bias lead configurations like in Fig. \ref{fig:diag}(b). 
However, the effect of such non-uniform current biasing  on SQUID array performance  is only  now being  studied. 
In this section we will compare the performance of uniformly biased arrays (Fig. \ref{fig:diag}(a)) with the performance of centre biased arrays (Fig. \ref{fig:diag}(b)).

\subsection{Parameters and normalisation}

\begin{figure}
    \centering
    \includegraphics[width=0.5\textwidth]{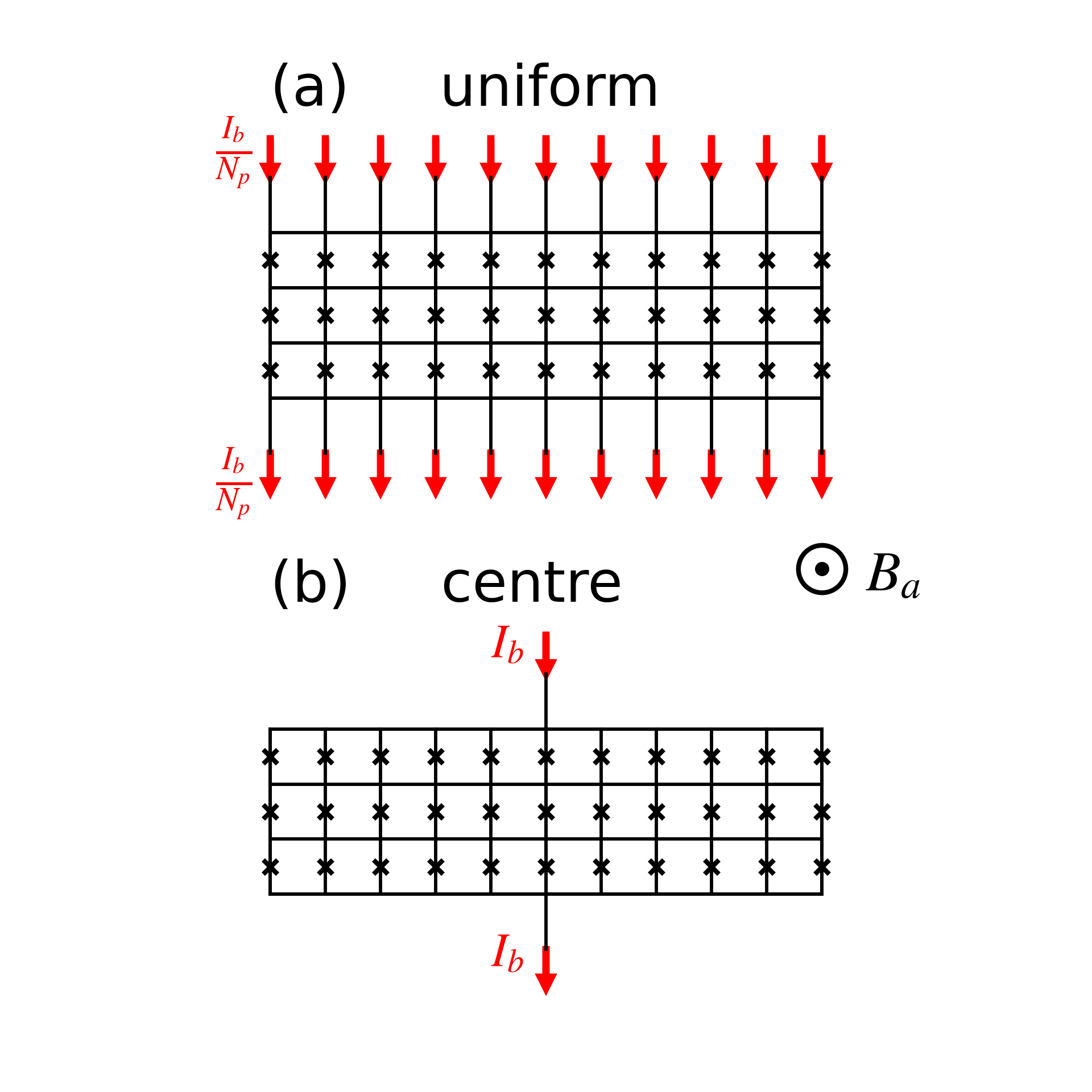}
    \caption{(a) uniform biased array and (b) centre biased array, with crosses depicting the Josephson junctions and red arrows the input and output bias currents of each configuration. $B_a$ is the applied magnetic field, $B_a=\Phi_a/A$ where $A$ is the SQUID cell area.}
    \label{fig:diag}
\end{figure}

In this paper we will use the usual normalisation for the array voltage, bias currents and magnetic flux \cite{Tesche1977, Gali2022a}. With $I_c$ and $R$ the critical current and normal resistance of a JJ,  we define the normalised time-averaged voltage $\bar{v}=\bar{V}/(RI_c)$ with $\bar{V}$ the time-averaged voltage between the top and bottom lead (Fig. \ref{fig:diag}); the normalised bias current $i_b=I_b/(N_p I_c)$ with $I_b$ the total bias current; and the normalised applied homogeneous magnetic flux $\phi_a = \Phi_a/\Phi_0$ where $\Phi_0$ is the flux quantum.
We define the voltage modulation depth as $\Delta \bar{v}=\max \left[ \bar{v}(\phi_a) \right] - \min \left[ \bar{v}(\phi_a) \right]$ and the transfer function as $\bar{v}_{\phi} = \partial \bar{v} / \partial \phi_a$.
To be as general as possible when characterising an array we use the dimensionless parameters $\beta_L=2L_s I_c/\Phi_0$ (screening parameter), $\Gamma= 2\pi k_B T / (I_c \Phi_0)$ (thermal noise strength) and $\kappa= L_s^k/L_s$ (kinetic self-inductance fraction).
Here $L_s$ is the self-inductance of a single cell (which includes the kinetic self-inductance), $k_B$ the Boltzmann constant, $T$ the operating temperature and $L_s^k$ the kinetic self-inductance of a single cell.
2D arrays are labelled $(N_s, N_p)$-arrays where $N_s$ is the number of JJs in series and $N_p$ the number of JJs in parallel. 1D parallel arrays have $N_s =1$. 
For this study we assume square cells, i.e. SQUID cells with equal height and width.
From $\bar{v}_{\phi}$ one can obtain the magnetic-field-to-voltage transfer function $\partial \bar{V} / \partial B_a= (RI_c A / \Phi_0 ) \bar{v}_{\phi}$ where $A$ is the area of a SQUID cell. 

\subsection{$\bar{v}(\phi_a)$, $\bar{v}_{\phi}(i_b)$ and $\Delta \bar{v} (i_b)$ for uniform and centre biasing} 

\begin{figure}[h!]
    \centering
    \includegraphics[width=0.5\textwidth]{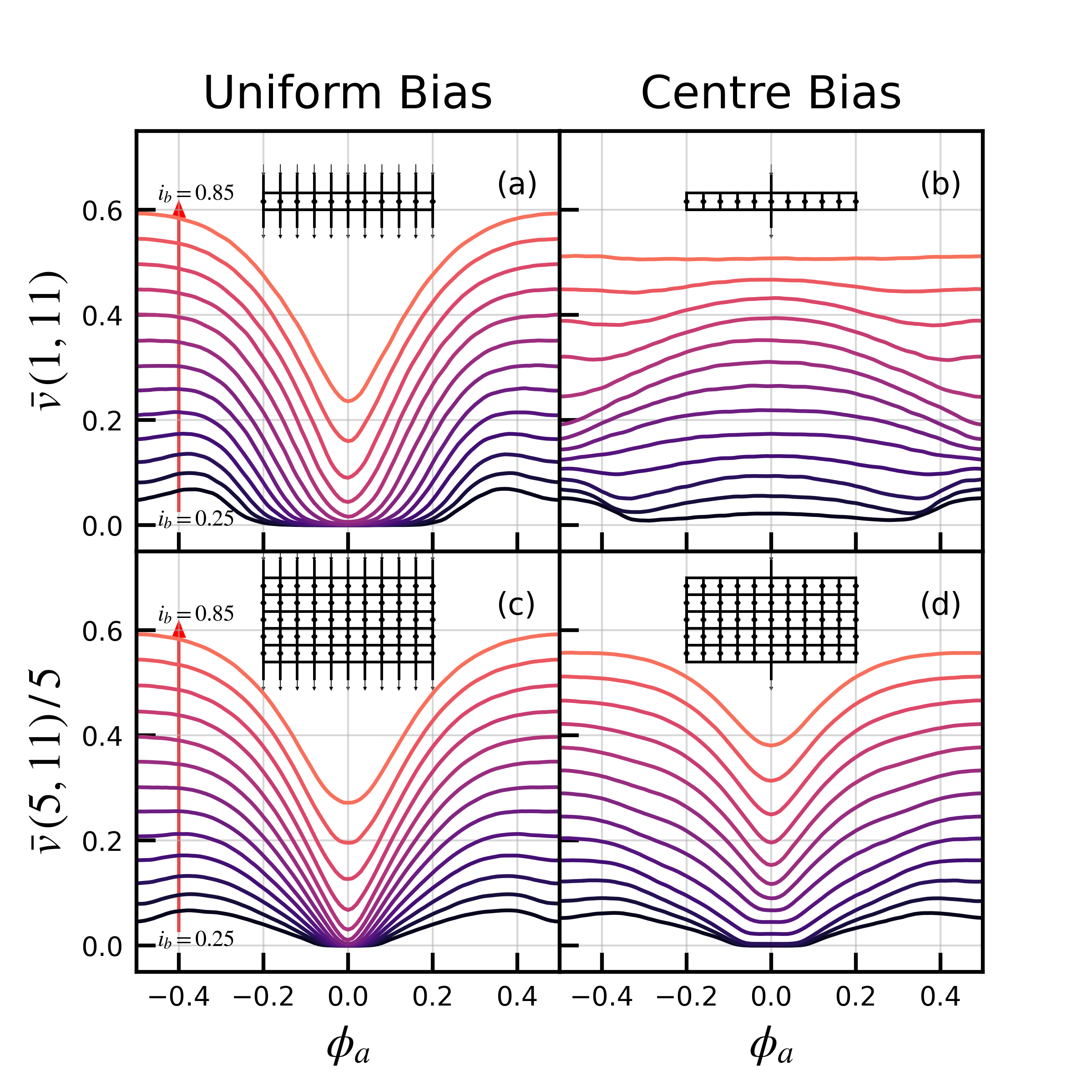}
    \caption{Comparing the effect of bias configurations on $\bar{v}/N_s$ versus $\phi_a$ using different bias currents $i_b=0.25,0.3,…,0.8,0.85$. (a) and (b) represent 1D SQUID arrays, and (c) and (d) represent (5,11)-SQUID arrays. For these arrays $\beta_L=1$, $\Gamma=0.16$ and $\kappa=0.5$. 
    }
    \label{fig:v-vs-phi}
\end{figure}

Figure \ref{fig:v-vs-phi} shows the normalized time-averaged voltage $\bar{v}(N_s,11)/N_s$ versus the normalised applied magnetic flux $\phi_a$ for a set of 1D ($N_s=1$) and 2D SQUID arrays and for a range of normalised bias currents $i_b \in [0.25,  0.85]$, here $\beta_L=1$, $\Gamma=0.16$ and $\kappa=0.5$. 
In Fig. \ref{fig:v-vs-phi}, $N_p= 11$ junctions in parallel is chosen in our analysis, since $N_p=11$ is large enough for the effects of centre biasing to become apparent.
To compare the effects of $N_s$, we use $N_s=1$ and $N_s=5$.
Figures \ref{fig:v-vs-phi}(a) and \ref{fig:v-vs-phi}(b) reveal that for $N_s=1$ (1D parallel arrays), the centre biased array shows reduced voltage modulation $\Delta \bar{v}$ and reduced maximum transfer function $\bar{v}_{\phi}^{\max}=\max \left[ \bar{v}_{\phi}(\phi_a) \right]$ compared with the uniformly biased array. 
In addition, for the 1D centre biased array, $\Delta \bar{v}$ shows a local maximum at $\phi_a=0$ (Fig. \ref{fig:v-vs-phi}(b)) in sharp contrast to the minimum usually present for dc-SQUID and SQUID arrays uniformly biased \cite{Tesche1977, Clarke2004, Mitchell2019, Gali2022a}.

Similar anomalies in the voltage response have been noted experimentally for 1D parallel arrays with large $N_p$ \cite{Mitchell2019}. 
Not only do the $\bar{v}(\phi_a)$ curves show smaller $\bar{v}_{\phi}^{\max}$ for the 1D centre biased arrays but, in the range $\phi_a \in [ 0, 0.5 ] $, $\bar{v}_{\phi}^{\max}$ also changes sign compared with the uniform bias case (Fig. \ref{fig:v-vs-phi}(a)). 
Figures \ref{fig:v-vs-phi}(c) and \ref{fig:v-vs-phi}(d) compare the voltage response of 2D SQUID arrays with $N_s=5$ for uniform and centre biasing respectively.
For the centre biased 2D arrays (Fig. \ref{fig:v-vs-phi}(d)), the anomalous behaviour seen for the centre biased 1D parallel array (Fig. \ref{fig:v-vs-phi}(b)) is no longer present.
However, there is a reduction in $\Delta \bar{v}$ and $\bar{v}_{\phi}^{\max}$ for nearly all $i_b$ compared to the uniformly biased array (Fig. \ref{fig:v-vs-phi}(c)).

\begin{figure}[h!]
    \centering
   \includegraphics[width=0.5\textwidth]{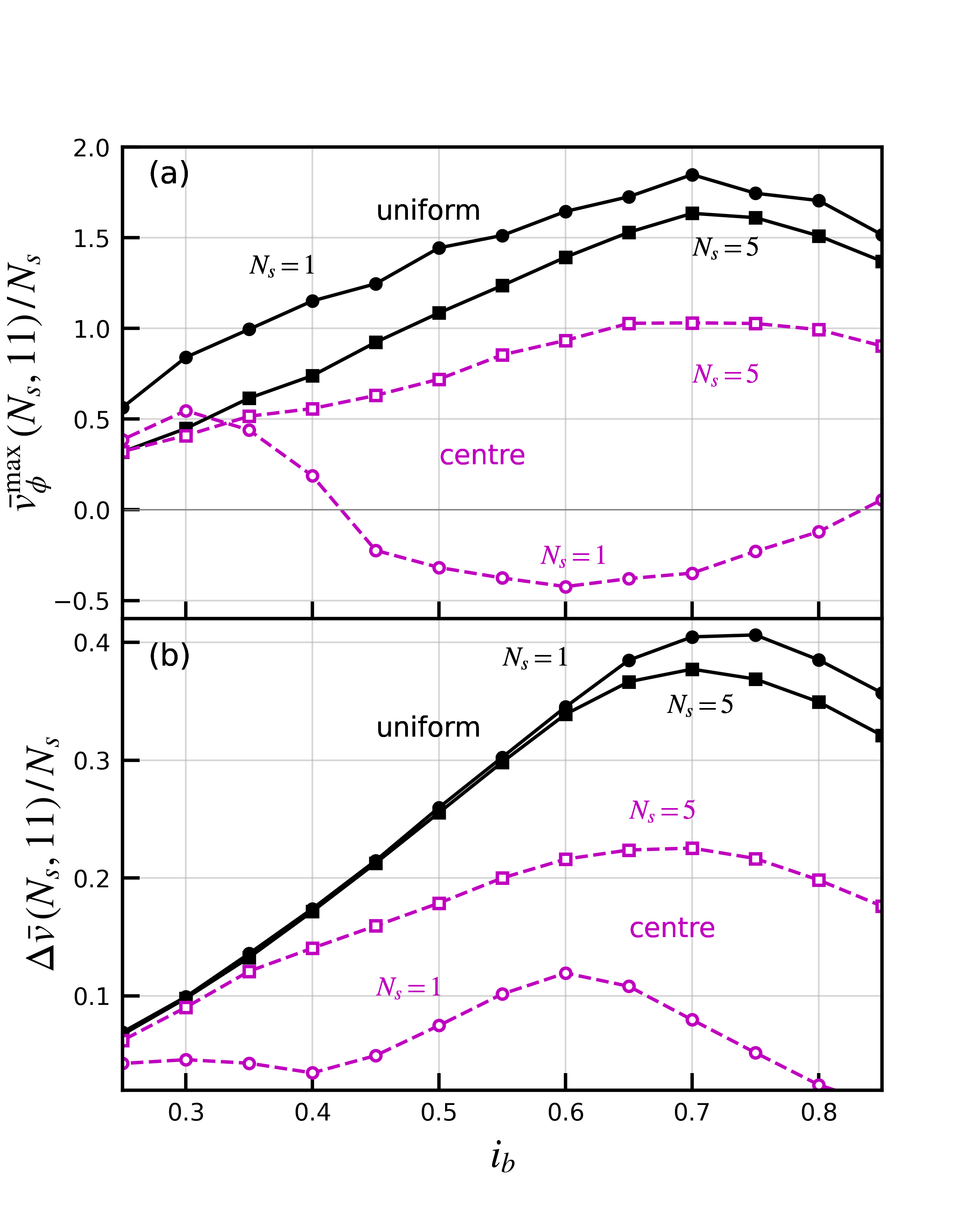}
   \caption{(a) $\bar{v}_{\phi}^{\max}/N_s$ and (b) $\Delta \bar{v}/N_s$  calculated in the half-period range $\phi_a \in [0, 0.5]$ versus $i_b$ of centre (magenta dashed lines with open symbols) and uniform (black solid lines with filled symbols) biased arrays for $N_s=1$ (circles) and $N_s=5$ (squares). For these arrays $\beta_L=1$, $\Gamma =0.16$ and $\kappa=0.5$. 
   }
    \label{fig:dv_vs_ib}
\end{figure}

Figure \ref{fig:dv_vs_ib} depicts (a) $\bar{v}_{\phi}^{\max}/N_s$ and (b) $\Delta \bar{v}/N_s$ versus $i_b$ calculated in the half-period range $\phi_a \in [0, 0.5]$ for the same arrays studied in Fig. \ref{fig:v-vs-phi}.
Here, it can be seen clearly that, for all bias currents and $N_s$, uniformly biased arrays perform better than centre biased ones.
For instance, for the 2D array biased at the optimal current $i_b^*$, the optimal transfer function $\bar{v}_{\phi}^*$ of the centre biased array is approximately 30\% smaller than the one of the uniform array (Fig. \ref{fig:dv_vs_ib}(a)), and the optimal voltage modulation depth $\Delta \bar{v}^*$ of the centre biased array is approximately half of the uniform one (Fig. \ref{fig:dv_vs_ib}(b)).  

Comparing the solid lines for the uniform biased arrays in Figs. \ref{fig:dv_vs_ib}(a) and \ref{fig:dv_vs_ib}(b), the optimal bias current $i_b^*$ values at which the $\bar{v}_{\phi}^{\max}$ curves optimize are very similar to the $i_b^*$ at which $\Delta \bar{v}$ optimizes, i.e., $i_b \approx 0.7 - 0.8$. 
A similar optimal bias current range occurs for the centre biased 2D array ($N_s =5$, open squares). The exception to this behaviour is for the centre biased 1D parallel arrays (magenta dashed lines with open circles), where the optimum $\Delta \bar{v}$ is at $i_b^*=0.6$ whilst $|\bar{v}_{\phi}^{\max}|$  shows two clear maxima, at $i_b=0.3$ ($\bar{v}_{\phi}^{\max} > 0$) and at $i_b = 0.6$ ($\bar{v}_{\phi}^{\max} < 0$).
This sign change in $\bar{v}_{\phi}^{\max}$ with $i_b$ is in agreement with previously reported experimental measurements (Fig. 3(a) in \cite{Mitchell2019}). 
To be consistent, throughout the rest of this paper, we will define the optimal bias current $i_b^*$ as the bias current that optimises $\Delta \bar{v}$, since it is well defined for both configurations.

\subsection{Non-uniform fluxoid distribution created by the centre bias configuration}

\begin{figure}[h!]
    \centering
    \includegraphics[width=0.5\textwidth]{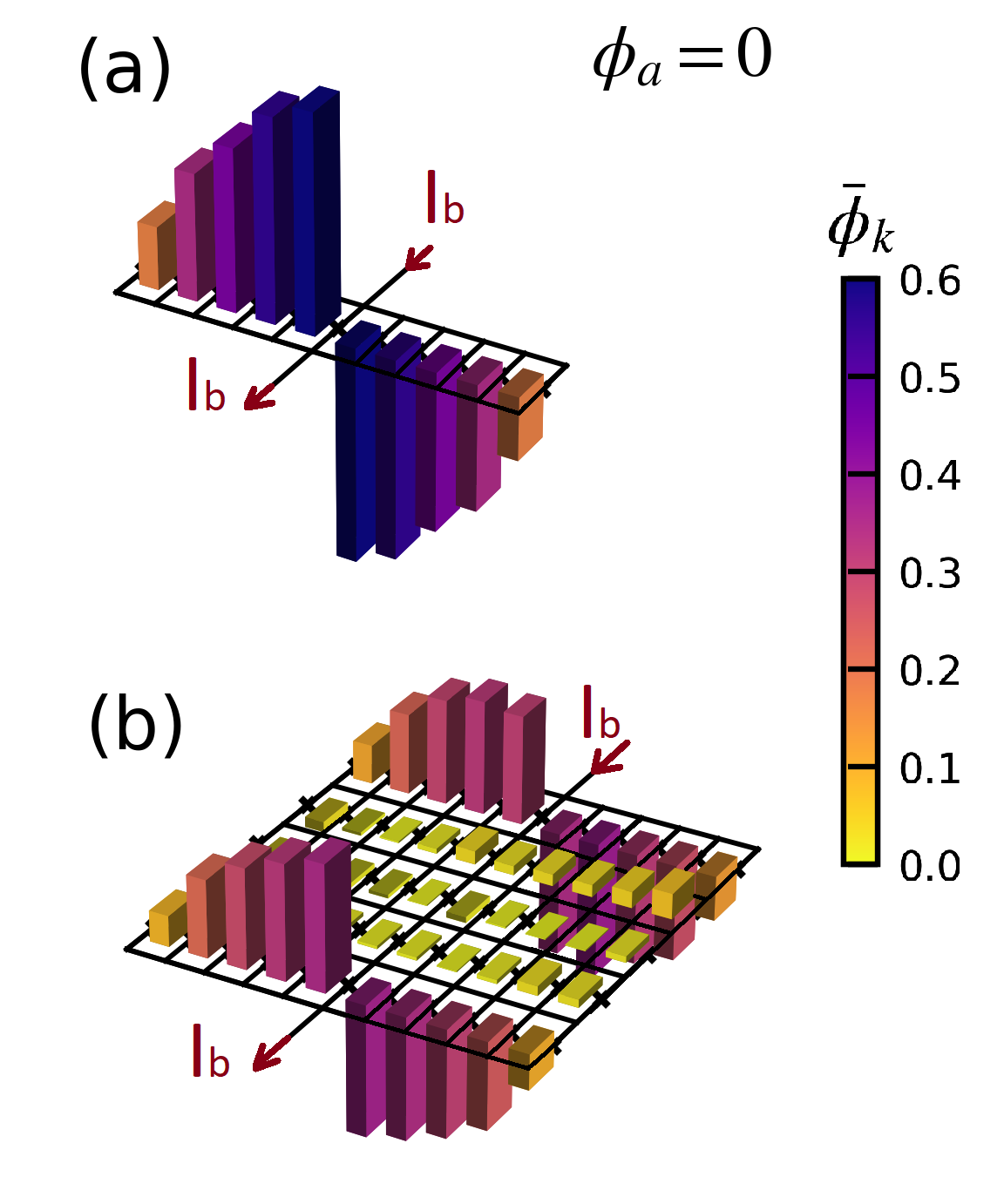}
    \caption{Time-averaged fluxoid $\bar{\phi}_k$ in each cell of centre biased  $(N_s,11)$-SQUID arrays with $\beta_L=1$, $\Gamma=0.16$ and $\kappa =0.5$ at $\phi_a=0$ and $i_b^*$. (a) 1D array and (b) 2D array with $N_s=5$.}
    \label{fig:3d-fluxes}
\end{figure}

Experimental results for 1D SQUID arrays have shown a decline in the optimal transfer function and deviations from the expected sinusoidal-like voltage versus magnetic field response for centre biased 1D parallel arrays as the number of junctions (and loops) in parallel increases \cite{Mitchell2019}. 
The decline in performance for centre biased 1D arrays was attributed to the different amount of fluxoids coupled into the various SQUID loops created by the larger bias currents forced to flow horizontally (Fig. \ref{fig:diag}(b)). 
Figure \ref{fig:3d-fluxes} shows a bar graph of the normalised time-averaged fluxoid per SQUID loop, $\bar{\phi}_k = \bar{\Phi}_k / \Phi_0$, for each loop in a centre biased 1D (Fig. \ref{fig:3d-fluxes}(a)) and 2D (Fig. \ref{fig:3d-fluxes}(b)) array. 
The time-dependent fluxoid $\Phi_k$ is defined as
\begin{equation}
    \Phi_k = \Phi_a + \Phi_{L,k} + \mu_0 \lambda_L^2 \oint \vec{j} \cdot \vec{dl},
\end{equation}
where $\Phi_{L,k}$ is the magnetic flux created by geometric inductances, $\mu_0$ is the vacuum permeability, $\lambda_L$ the London penetration depth of the device material and $\vec{j}$ is the supercurrent density. 
The line integration is defined anticlockwise around a cell.
In Fig. \ref{fig:3d-fluxes}, $N_p = 11$, $\beta_L=1$, $\Gamma=0.16$, $\kappa=0.5$ and $\phi_a=0$. 
For uniformly biased arrays the fluxoids are negligibly small compared to the fluxoids generated in centre biased arrays, and therefore no figures are shown.

Figure \ref{fig:3d-fluxes}(a) shows that for a centre biased $(1,11)$-SQUID array the fluxoids are largest in the two cells near the array centre and decay anti-symmetrically with respect to the centre bias leads. 
For the $N_p$, $\beta_L$, $\Gamma$ and $\kappa$ used here, the maximum fluxoid reaches a magnitude of $\sim 0.6 \Phi_0$  at the central loops of the array.
This large and non-uniform fluxoid distribution across 1D centre biased arrays explains the unusual voltage to flux response seen in Fig. \ref{fig:v-vs-phi}(b), with a voltage maximum at $\phi_a=0$ and reduced transfer-function compared with the uniformly biased equivalent array. 
Figure \ref{fig:3d-fluxes}(b) highlights that the first and last row respond to the centre bias current in a similar way as the 1D array although now the fluxoid values are reduced by about a factor of 2.
The middle rows of the array (rows 2, 3 and 4 in this $N_s =5$ array example) have much lower fluxoid values because the first and last row redistribute the biasing current more uniformly across the full width of the array. 
This redistribution of the central bias current across the array by the first and last row explains why the voltage responses of the centre biased 2D array in Fig. \ref{fig:v-vs-phi}(d) show similar shapes as the voltage responses of the equivalent uniformly biased array in Fig. \ref{fig:v-vs-phi}(c). 

\subsection{Optimal transfer function dependence on $N_p$, $N_s$, $\beta_L$ and $\Gamma$ and biasing scheme comparison}

The optimal transfer function $\bar{v}_{\phi}^*= \bar{v}_{\phi} (i_b^*, \phi_a^* )$ is the transfer function evaluated at the optimal applied magnetic flux $\phi_a^*$ and biased at the optimal bias current $i_b^*$. 
In this section, we compare $\bar{v}_{\phi}^*$ of uniform and centre bias configurations depending on $N_s$ and $N_p$ as a function of $\beta_L$ and $\Gamma$ at $\kappa=0.5$.

\begin{figure}
    \centering
    \hspace*{-3mm}
    \includegraphics[width=0.5\textwidth]{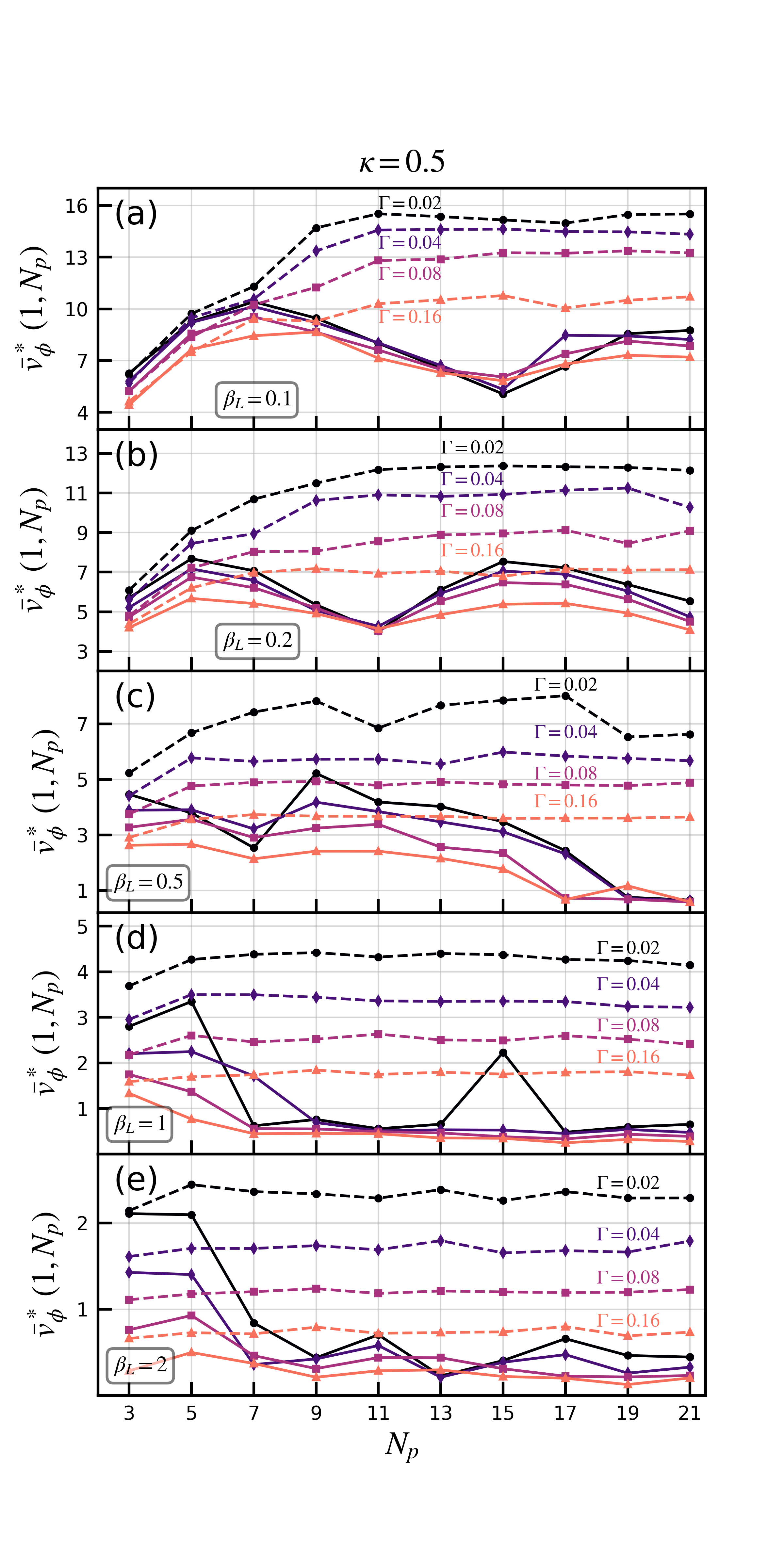}
    \caption{Optimal transfer function of 1D arrays versus $N_p$ for centre bias (solid lines) and uniform bias (dashed lines) for different screening parameters: (a) $\beta_L=0.1$, (b), $\beta_L=0.2$, (c) $\beta_L=0.5$, (d) $\beta_L=1$ and (e) $\beta_L=2$. For each $\beta_L$ the kinetic self-inductance fraction is fixed, $\kappa=0.5$, and different thermal noise strengths are shown: $\Gamma=0.02$ (black lines with circles), $\Gamma=0.04$ (purple lines with diamonds), $\Gamma=0.08$ (magenta lines with squares), $\Gamma=0.16$ (orange lines with triangles). }
    \label{fig:dvdp_opt-vs-Np}
\end{figure}

Figure \ref{fig:dvdp_opt-vs-Np} shows the optimal transfer function of 1D parallel SQUID arrays with $N_p=11$, i.e. $(1, 11)$-arrays, centre biased (solid lines) and uniformly biased (dashed lines) for different screening parameters (a) $\beta_L=0.1$, (b) $\beta_L=0.2$, (c) $\beta_L=0.5$, (d) $\beta_L=1$ and (e) $\beta_L=2$. For each $\beta_L$ we show four different noise thermal strengths, $\Gamma=0.02$ (black lines with circles), $\Gamma=0.04$ (purple lines with diamonds), $\Gamma=0.08$ (magenta lines with squares) and $\Gamma=0.16$ (orange lines with triangles).

The centre biased arrays (solid lines) present pronounced oscillations in $\bar{v}_{\phi}^*$  with $N_p$ which have not been reported previously (note that y-axis have different scales). 
These oscillations are due to the large and unequal fluxoid values per cell created by the bias current distribution (see Fig. \ref{fig:3d-fluxes}) and are amplified due to the bias current optimisation. 
The $\bar{v}_{\phi}^* (1, N_p)$ minima depend on the total fluxoid in half the array. For instance, the clearly visible minimum in Figs. \ref{fig:dvdp_opt-vs-Np}(a) ($N_p=15$) and \ref{fig:dvdp_opt-vs-Np}(b) ($N_p=11$), appear when the sum of the fluxoids in the cells on the left half of the array is $\sim \Phi_0$, and the next minimum appears when the sum is $\sim 2\Phi_0$.
Considering that the total bias current applied to a centre biased array increases with $N_p$ according to $I_b=i_b^* N_p I_c$,   large $\beta_L$’s will create large fluxoid values even for small $N_p$, while small $\beta_L$’s require larger bias currents, and therefore larger $N_p$, to generate a full $\Phi_0$. 
This fluxoid dependence on $N_p$ and $\beta_L$ is responsible for the $N_p$-oscillations of $\bar{v}_{\phi}^*$ for 1D arrays (Fig. \ref{fig:dvdp_opt-vs-Np}) and the overall decrease of $\bar{v}_{\phi}^*$ with $N_p$.

The uniformly biased arrays (dashed lines) show the same plateauing behaviour previously reported for $\bar{v}_{\phi}^{\max}$ \cite{Mitchell2019, Gali2022b, Gali2022a}. This trend depends on the coupling radius which describes the number of junctions that interact or couple with each other \cite{Kornev2009, Kornev2011, Gali2022b}.
Small $N_p$-oscillations also appear for the uniform case, which are more noticeable for small $\beta_L$ and $\Gamma$.

Figure \ref{fig:dvdp_opt-vs-Np} shows that decreasing $\Gamma$ increases the transfer function of uniform biased arrays, while for centre biased arrays decreasing $\Gamma$ increases the amplitude of the $N_p$-oscillations of $\bar{v}_{\phi}^*$.
Figures \ref{fig:dvdp_opt-vs-Np}(d) and (e) show that for small $\Gamma$'s some peaks appear at certain $N_p$. 
These smaller oscillations are more obvious for $\beta_L=2$, appearing at $N_p=11$ and 17, because these arrays can store more magnetic energy, these oscillations are more resilient to thermal noise.

\begin{figure}
    \centering
    \includegraphics[width=0.5\textwidth]{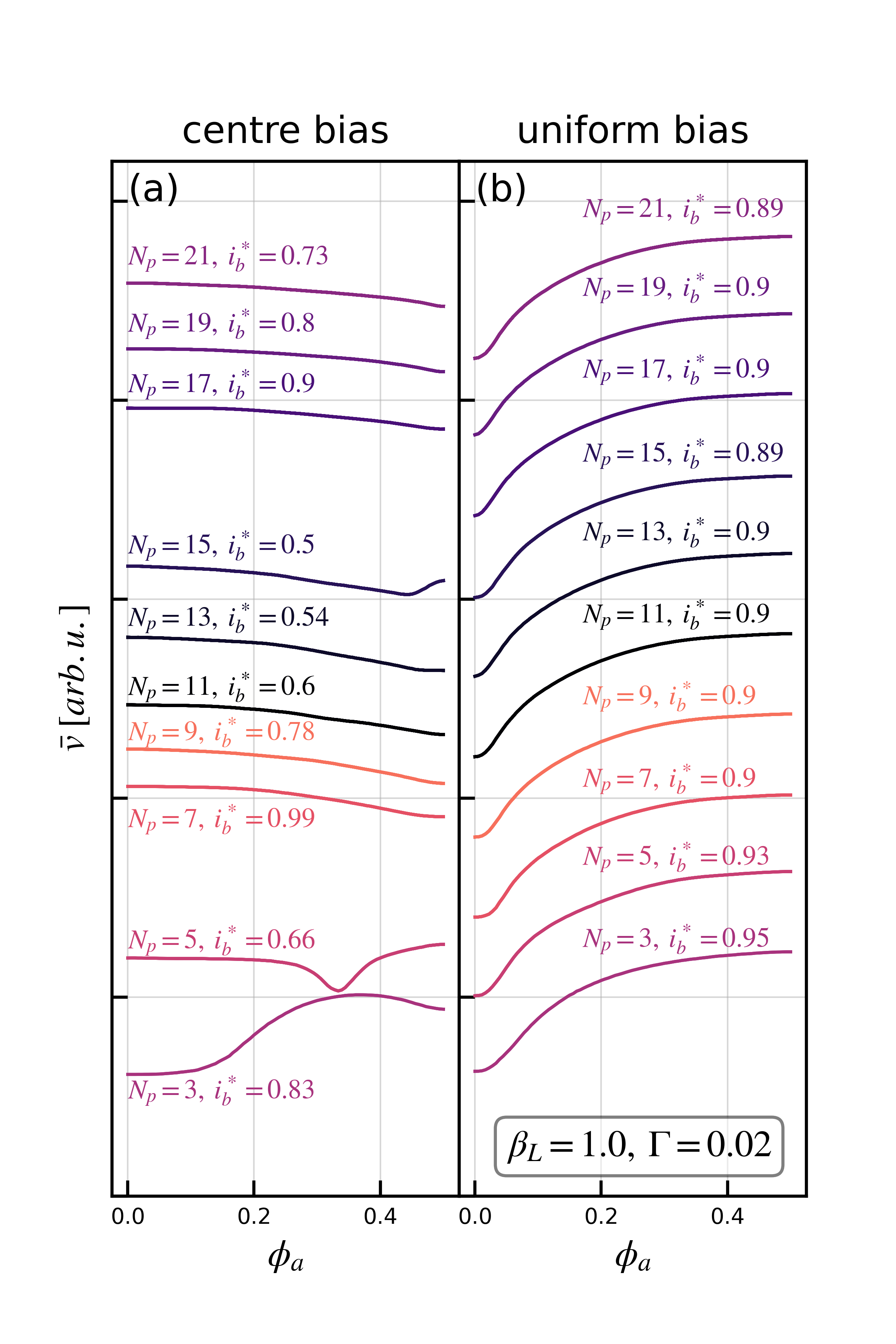}
    \caption{$\bar{v} \, (1, N_p)$ versus $\phi_a$ responses in arbitrary units for (a) centre and (b) uniform biased 1D parallel SQUID arrays with $\beta_L=1$, $\Gamma=0.02$ and $\kappa=0.5$ for different $N_p$ biased at $i_b^*$. For clarity, each $\bar{v}(\phi_a)$ response has been offset.}
    \label{fig:v-vs-phi_Np}
\end{figure}

Figure \ref{fig:v-vs-phi_Np} depicts the flux-to-voltage response $\bar{v}(\phi_a)$ in arbitrary units for (a) centre and (b) uniformly biased 1D SQUID arrays with $\beta_L=1$, $\Gamma = 0.02$ and $\kappa=0.5$ for different $N_p$'s at $i_b^*$.
For clarity, the $\bar{v}(\phi_a)$ responses have been offset for each $N_p$.
Figure \ref{fig:v-vs-phi_Np}(a) demonstrates that $i_b^*$ greatly varies with $N_p$. For $N_p=3$ the degrading effect of the centre bias configuration is still weak and the $\bar{v}(\phi_a)$ response and $i_b^*$ are similar to the uniformly biased case.
For $N_p \geq 5$ the voltage responses for centre biasing look very different to the ones of uniform biasing with $\Delta\bar{v}$ being very small, which explains the quick decrease of $\bar{v}_{\phi}^*$ with $N_p$.
On the other hand, the voltage response and $i_b^*$ of uniformly biased arrays show very small variations with $N_p$ which is reflected by the negligible $N_p$-oscillations in Fig. \ref{fig:dvdp_opt-vs-Np}.

\begin{figure}[h!]
    \centering
    \includegraphics[width=0.5\textwidth]{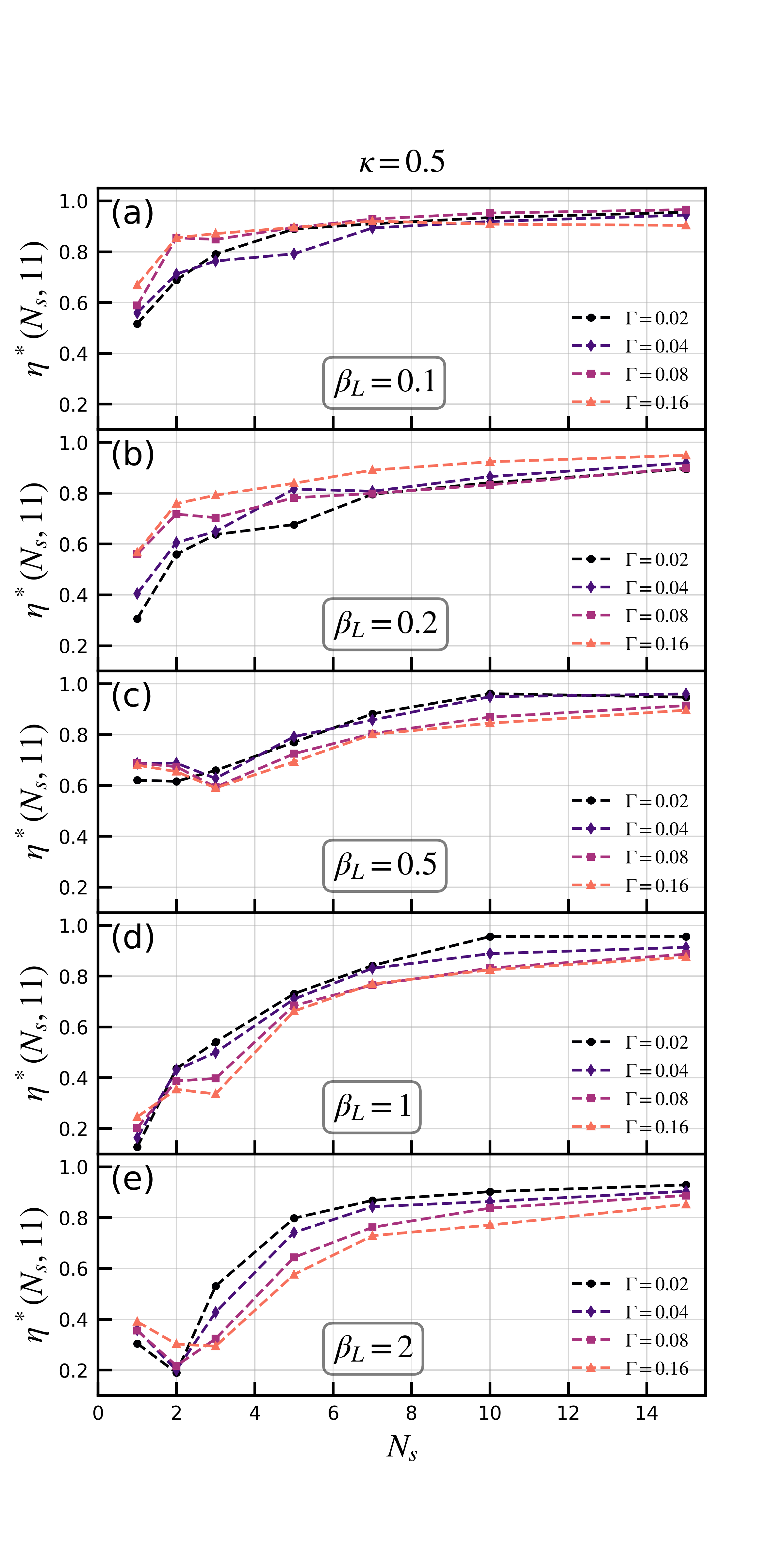}
    \caption{$\eta^* (N_s, 11)$ versus $N_s$ for different screening parameters: (a) $\beta_L=0.1$, (b) $\beta_L=0.2$, (c) $\beta_L=0.5$, (d) $\beta_L=1$ and (e) $\beta_L=2$. For each $\beta_L$ the kinetic self-inductance fraction is fixed, $\kappa=0.5$, and different thermal noise strengths are shown: $\Gamma=0.02$ (black dashed line with circles), $\Gamma=0.04$ (purple dashed line with diamonds), $\Gamma=0.08$ (magenta dashed line with squares), $\Gamma=0.16$ (orange dashed line with triangles).}
    \label{fig:dvdp_opt-vs-Ns}
\end{figure}

Next we investigate the optimal transfer function of 2D SQUID arrays. We compare $\bar{v}_{\phi}^*$ for centre biasing with $\bar{v}_{\phi}^*$ for uniform biasing by introducing the quantity $\eta^*$ as
\begin{equation}
\eta^* (N_s,N_p) = \frac{\bar{v}_{\phi}^* (N_s,N_p )|_{cent} }{ \bar{v}_{\phi}^* (N_s,N_p ) |_{unif} }.
\end{equation}

Thus, for a given $N_p$, $\beta_L$, $\Gamma$ and $\kappa$, this ratio will reveal what $N_s$ is needed for the performance of a centre biased 2D array to approach that of a uniformly biased one, i.e. for  $\eta^*$ to approach 1. 
Figure \ref{fig:dvdp_opt-vs-Ns} depicts $\eta^* (N_s,11)$ as a function of $N_s$ for several screening parameters: (a) $\beta_L=0.1$, (b) $\beta_L = 0.2$, (c) $\beta_L=0.5$, (d) $\beta_L=1$ and (e) $\beta_L=2$. For each $\beta_L$ we analyse different thermal noise strengths: $\Gamma=0.02$ (black dashed line with circles), $\Gamma=0.04$ (purple dashed line with diamonds), $\Gamma=0.08$ (magenta dashed line with squares), $\Gamma=0.16$ (orange dashed line with triangles).
The response of $\eta^*$ on $N_s$ can be split into two regimes; $N_s<3$, where the 1D behaviour dominates, and $N_s \geq 3$, where the 2D behaviour takes over.  

The fluxoid distribution shown in Fig. \ref{fig:3d-fluxes}(b) shows that the first and last row are most affected by the bias configuration, while the middle rows perceive a uniform distribution of the bias current, which explain the $\eta^* (N_s)$ dependence and distinction between $N_s<3$ and $N_s \geq 3$.  

Figure \ref{fig:dvdp_opt-vs-Ns} reveals the detrimental effect that the centre bias configuration has as $\beta_L$ increases. 
The effect of $\Gamma$ on $\eta^*$ is relatively small with the  $\eta^*$ versus $N_s$ trends mainly determined by $\beta_L$.
As $\beta_L$ increases $N_s$ must increase to achieve $\eta^* > 0.8$ and also the effects of $\Gamma$ become more important.
For the larger $\beta_L=1$ and 2, smaller $\Gamma$ gives slightly higher $\eta^*$, while for smaller $\beta_L$ there is no clear dependence.

Studying the dependence on $N_s$, one sees that for $N_s \geq 3$ the 2D effects dominate and $\eta^*$ increases with $N_s$.
For $N_s/(N_p-1) \geq 1$ then $\eta^* \gtrsim 0.8$ for all $\beta_L$ and $\Gamma$. 
This gives an estimate of the minimum $N_s$ needed for a given $N_p$, so that the centre biased array reaches the performance of at least  80\% of the uniformly biased array.
     
Finally, it is important to note that the results shown in this work assume square cells, i.e. SQUID cells where height and width are equal. If one would investigate SQUID arrays where the SQUID cell width is larger than the height, then the detrimental effect of the centre bias configuration would be even stronger and the $N_p$-oscillations shown in Fig. \ref{fig:dvdp_opt-vs-Np} would appear at smaller $N_p$ values while larger $N_s$ would be needed for $\eta^*>0.8$. 
On the other hand, arrays where the SQUID cell width is smaller than the height would perform better.

\subsection{Optimal transfer function dependence on $\kappa$}

In the previous sections we have investigated for $(N_s, N_p)$-arrays the dependence of $\bar{v}_{\phi}^*$ on the parameters $\beta_L$ and $\Gamma$, and have kept $\kappa$ constant at $\kappa=0.5$.
The value of $\kappa$ is affected by $\lambda_L$, the London penetration depth of the superconducting material the array is made from.
Specifically, $\kappa = L_s^k / L_s$ with the kinetic self-inductance $L_s^k$ being proportional to $\lambda_L^2$.

To investigate the effect of $\kappa$ on $\bar{v}_{\phi}^*$ and thus $\eta^*$, we choose a representative value for the thermal noise strength of $\Gamma=0.08$ and scan over the same $(N_s, N_p, \beta_L)$ parameter space as we did above.
Here we study three scenarios: the cell inductance is purely geometric ($\kappa=0$), the kinetic and geometric inductances are equal ($\kappa = 0.5 $), and the kinetic inductance dominates ($\kappa =0.9$).
The case $\kappa=0$ corresponds to a LTS device, while $\kappa=0.5$ corresponds to a HTS device \cite{Keenan2021, Muller2021}.

\begin{figure}[!h]
    \centering
    \includegraphics[width=0.5\textwidth]{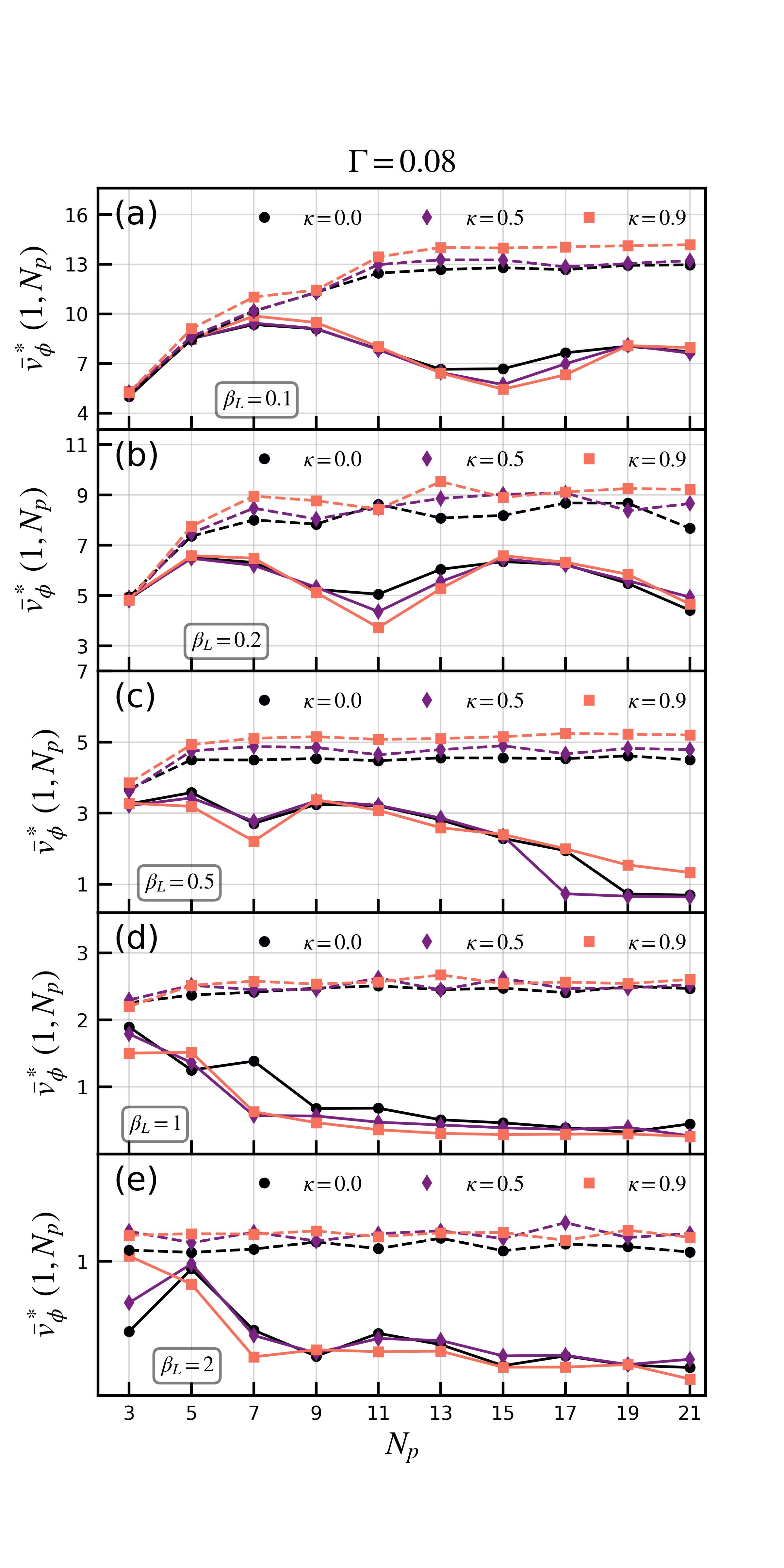}
    \caption{
    Optimal transfer function of 1D arrays versus $N_p$ for centre bias (solid lines) and uniform bias (dashed lines) with $\Gamma=0.08$ and for different screening parameters: (a) $\beta_L=0.1$, (b), $\beta_L=0.2$, (c) $\beta_L=0.5$, (d) $\beta_L=1$ and (e) $\beta_L=2$. For each $\beta_L$ different $\kappa$ are shown: $\kappa=0$ (black lines with circles), $\kappa=0.5$ (purple lines with diamonds) and $\kappa=0.9$ (orange lines with squares).
    }
    \label{fig:dvdp-vs-Np_kinetic}
\end{figure}

\begin{figure}[!h]
    \centering
    \includegraphics[width=0.5\textwidth]{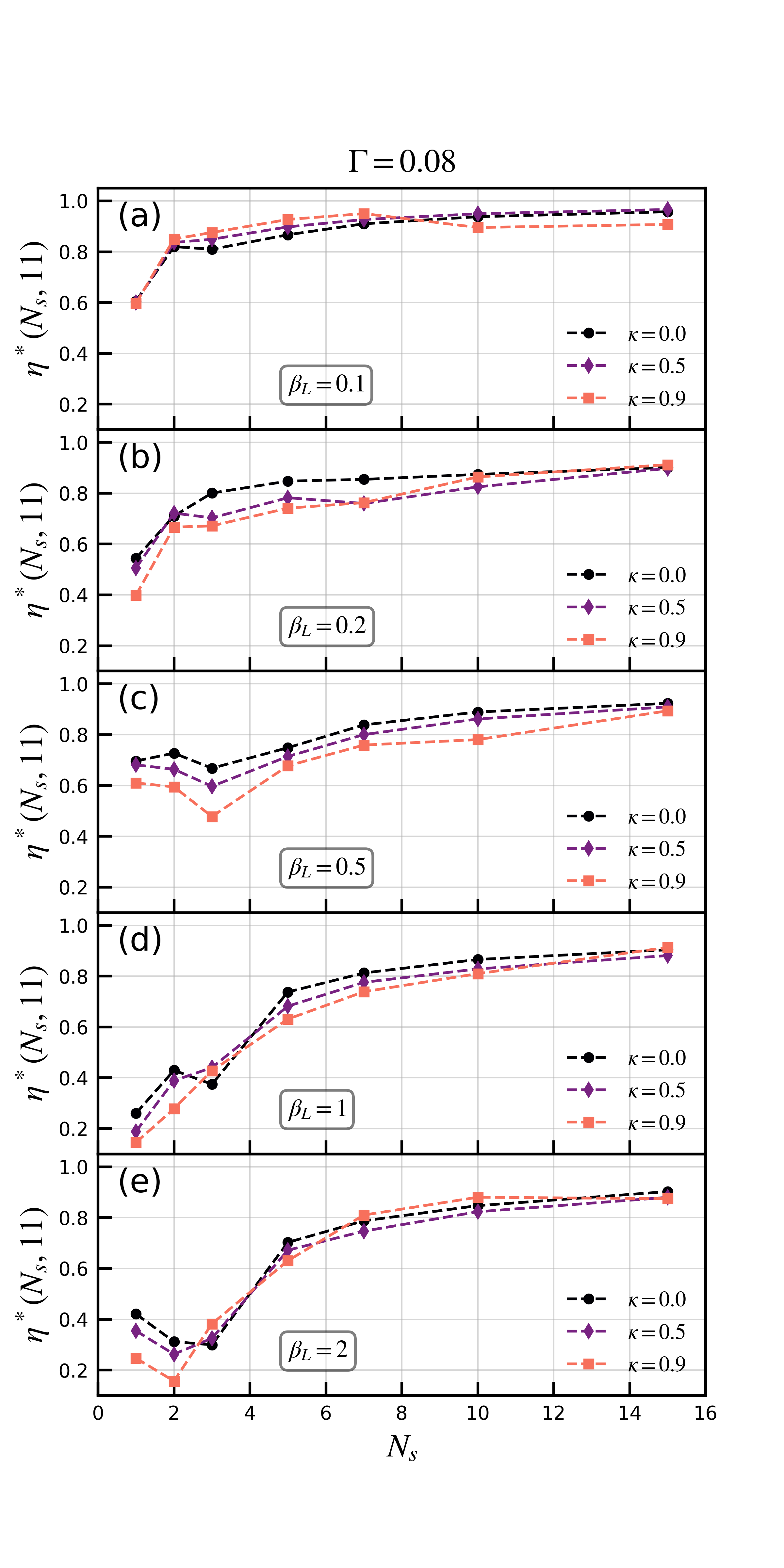}
    \caption{
    $\eta^* (N_s, 11)$ versus $N_s$ with $\Gamma=0.08$ and for different screening parameters: (a) $\beta_L=0.1$, (b) $\beta_L=0.2$, (c) $\beta_L=0.5$, (d) $\beta_L=1$ and (e) $\beta_L=2$. For each $\beta_L$ different $\kappa$ are shown: $\kappa=0$ (black dashed lines with circles), $\kappa=0.5$ (purple dashed lines with diamonds) and $\kappa=0.9$ (orange dashed lines with squares).}
    \label{fig:eta-vs-Ns_kinetic}
\end{figure}

Figure \ref{fig:dvdp-vs-Np_kinetic} shows the optimal transfer function $\bar{v}_{\phi}^*$ versus $N_p$ for different screening parameters (a) $\beta_L=0.1$, (b) $\beta_L = 0.2$, (c) $\beta_L=0.5$, (d) $\beta_L=1$ and (e) $\beta_L=2$. For each $\beta_L$ we analyse different kinetic self-inductance fractions: $\kappa=0$ (lines with black circles), $\kappa=0.5$ (lines with purple diamonds) and $\kappa=0.9$ (lines with orange squares).
For centre biased arrays (solid lines), larger $\kappa$ produce slightly larger $N_p$-oscillations. This can be seen in Figs. \ref{fig:dvdp-vs-Np_kinetic}(a) and (b) where the only variations with $\kappa$ appear at the local minima, $N_p=15$ in Fig. \ref{fig:dvdp-vs-Np_kinetic}(a) and $N_p=11$ in Fig. \ref{fig:dvdp-vs-Np_kinetic}(b).
For uniform biased arrays (dashed lines), there is a slight increase of the optimal transfer function with increasing $\kappa$.

Figure \ref{fig:eta-vs-Ns_kinetic} presents the $\eta^*$ dependence on $N_s$ for different screening parameters (a) $\beta_L=0.1$, (b) $\beta_L = 0.2$, (c) $\beta_L=0.5$, (d) $\beta_L=1$ and (e) $\beta_L=2$. For each $\beta_L$ different kinetic self-inductance fractions are shown: $\kappa=0$ (dashed lines with black circles), $\kappa=0.5$ (dashed lines with purple diamonds) and $\kappa=0.9$ (dashed lines with orange squares).
Overall the trend of $\eta^*$ is dictated by $\beta_L$ with negligible effects due to $\kappa$ for $N_s > 3$.
For $N_s < 3$, $\eta^*$ decreases with increasing $\kappa$, which is partially due to the small increase of the optimal transfer function of uniformly biased arrays with $\kappa$.


\section{Summary}

It has been overlooked in the literature that the bias current configuration in 2D SQUID arrays can strongly affect the device flux-to-voltage response.
In this work we have generalised our previous theoretical model to now simulate arrays with different bias configurations. 
Using our model, researchers can estimate what bias configuration is more suitable for their design when uniform bias configuration is not possible. 

To investigate the importance of the bias current configuration, we have compared the uniform bias configuration with the centre bias one.
The two bias configurations show similar $\bar{v}$ vs $\phi_a$ curves for 2D arrays when $N_s \gtrsim N_p $, while centre biased 1D parallel arrays ($N_s=1$) present an unusual response with a local maximum at $\phi_a=0$. 
The optimal bias current $i_b^*$ which optimises the voltage modulation depth $\Delta \bar{v}$ and the maximum transfer function were found to depend on the array size $(N_s, N_p)$ and bias configuration. 
To understand the behaviours of 1D and 2D centre biased arrays, we have calculated the time-averaged fluxoid per SQUID cell.
Our simulations show that centre biasing creates a non-uniform fluxoid distribution in the first and last row of 2D SQUID arrays. 
These fluxoids created by the bias current account for the unusual voltage response observed.

The optimal transfer function $\bar{v}_{\phi}^*$ of uniformly and centre biased arrays has been investigated for arrays with different sizes $(N_s,N_p)$, for different screening parameters and thermal noise strengths, and for different kinetic self-inductance fractions $\kappa$. 
We have shown that for centre biased 1D parallel arrays $\bar{v}_{\phi}^*$ oscillates with $N_p$. 
Arrays with smaller $\beta_L$ show larger $\bar{v}_{\phi}^*$ and stronger $N_p$-oscillations.
Using $\eta^*=\bar{v}_{\phi}^*|_{cent} \, / \, \bar{v}_{\phi}^*|_{unif}$, our simulations indicate that the detrimental effect of a non-uniform bias configuration can be overcome by having arrays with at least the same number of SQUIDs in series as in parallel. 
Also, arrays with small $\beta_L$ are less affected by the bias configuration choice, and their optimal transfer function increases quicker with $N_s$.
We have shown that decreasing the thermal noise strength $\Gamma $ clearly increases the optimal transfer function of uniformly biased arrays.
While smaller $\Gamma$ enhance $\bar{v}_{\phi}^*$ for uniform biasing, for centre biased arrays this effect of $\Gamma$ is less pronounced.
Finally, we have investigated the role of the kinetic self-inductance, using the $\kappa$ parameter.
We found that $\bar{v}_{\phi}^*$ and $\eta^*$ are not strongly affected by $\kappa$.

\appendix

\section{Mathematical Framework}

In this section we extend our model for 2D SQUID arrays introduced previously \cite{Gali2022a} in order to include the effects of different bias configurations.
We define the input $\vec{I}_b$ and output $\vec{I}_f$ bias current vectors more generally as

\begin{align}
    \vec{I}_b &= (I_{b_1}, I_{b_2},\dots, I_{b_{N_p}}, 0, \dots, 0)^T, \\
    \vec{I}_f &= (I_{f_1}, I_{f_2},\dots, I_{f_{N_p}})^T,
    \label{eq:current-vectors}
\end{align}{}
where the superscript $T$ means transposition. Note that the only constraint for the input and output bias currents is that they must satisfy the conservation of current law, i.e. $I_b=\sum_{k=1}I_{b_k} = \sum_{k=1}I_{f_k}$.
This definition of $\vec{I}_b$ and $\vec{I}_f$ allows us to determine the amount of current at each bias lead, thus we can study any bias configuration where the leads are aligned with the junctions. For instance, to model the centre bias configuration represented in Fig. \ref{fig:diag}(b) all the vectors elements of $\vec{I}_b$ and $\vec{I}_f$ will be zero except for the elements corresponding to the middle junction lead, i.e. $I_{b_k} = I_{f_k} = I_b$ with $k=(N_p - 1)/2 + 1$ and $I_{b_j} = I_{f_j} = 0$ $\forall j\neq k$.

The different bias lead configurations affect the induced fluxes in the array. 
To account for the new bias lead inductances one has to replace the term $\hat{L}_b \vec{I}_b$ that appears in $\vec{\Phi}_{nf}$ below Eq. (25) in ref. \cite{Gali2022a} by $\hat{L}_{in} \vec{I}_b + \hat{L}_{out} \vec{I}_f $, where $\hat{L}_{in}$ ($\hat{L}_{out}$) describes the mutual inductance between each input (output) bias lead and each SQUID cell in the array.

\bibliography{biasconfig.bib}
\end{document}